\documentclass[reprint, showpacs, superscriptaddress, aps, pre, twocolumn, 10pt]{revtex4-1}

\usepackage{mathptmx}       
\usepackage{helvet}         
\usepackage{courier}        
\usepackage{type1cm}        
\usepackage{makeidx}         
\usepackage{graphicx}        
\usepackage[bottom]{footmisc}
\newcommand{\la}{\left\langle}
\newcommand{\ra}{\right\rangle}
\newcommand{\be}{\begin{equation}}
\newcommand{\ee}{\end{equation}}
\newcommand{\bse}{\begin{subequations}}
\newcommand{\ese}{\end{subequations}}
\newcommand{\bea}{\begin{eqnarray}}
\newcommand{\eea}{\end{eqnarray}}

\makeindex             

\begin{document}

\title{Hierarchical financial structures with money cascade}
\author{Mahendra K. Verma}
\email{mkv@iitk.ac.in}
\affiliation {Department of Physics, Indian Institute of Technology, Kanpur, India 208016}
%
%

\begin{abstract}
In this paper we show similarities between turbulence  and financial systems.  Motivated by similarities between the two systems, we construct a multiscale model for hierarchical financial structures that exhibits a constant cascade of wealth from large financial entities to small financial entities.  According to our model, large and intermediate scale financial institutions have a power law distribution. However, the wealth distribution is Maxwellian at individual scales.   
\end{abstract}

\maketitle

\section{Introduction}
\label{sec:mkv:intro}

A financial system is quite complex due to its multiscale and time-dependent nature.  Its complexity is accentuated by features such as saving, banking, corruption, stock-market, natural calamities, etc.  Despite its complicated structure, scientists have attempted to model financial systems using simple ideas.   One of the leading questions in this field is how to model the wealth and income distributions of individuals and companies~\cite{Chakrabarti:book:Econophysics}.   In this paper we will address this question.

Earlier models of income distribution of individuals are motivated by equilibrium statistical mechanics.  In such models, the individuals are mapped to  particles in a theromdynamic system, and economic activities to scattering among particles.  Following this analogy, it is expected that the income distribution follows Maxwellian or Gibbs distribution, similar to the distribution of kinetic energy in a gas container.  

The above distribution however holds only for low income groups. Pareto~\cite{Pareto:book} and others showed that  the individuals in a large income group exhibit power law distribution.  There have been many attempts to model this power law distribution using nonequilibrium nature of the system.  See Chakrabarti et al.~\cite{Chakrabarti:book:Econophysics} and references there in. 

In a financial system, wealth cascades from large financial entities  to smaller ones.  This cascade is somewhat similar to the cascade of kinetic energy in a turbulent system. In addition, a well-developed financial system contains income groups whose wealth has a wide range of distribution.  Also, note that these income groups interact with each other.  Motivated by the above similarities between turbulence and finance,  we construct a  model for a hierarchical financial system which is quite similar to   Kolmogorov's model for turbulent flow. 

The structure of the paper is as follows: In Section 2, we describe  a generic equilibrium model of a financial system. Section 3 contains a brief description of Kolmogorov's model for turbulence.  In Section 4 we construct a model for hierarchical financial system; this model is analogous to the Kolmogorov's model of turbulence.  We conclude in Section 5.

\section{Equilibrium model}
\label{sec:mkv:equilibrium}

In this section, we describe an equilibrium model of wealth distribution~\cite{Saha:book:Heat}. Before that we discuss  thermodynamics of an isolated gas reservoir in which  gas molecules interact with each other via collisions.  Under thermodynamic approximation,  all the molecules in the gas have approximate equal energy.  The variation in the energy of the molecules is given by Maxwell or Gibbs distribution~\cite{Landau:book:StatMech}:
\be
P(E) = \exp(-E/k_B T)
\ee
where $E=m v^2/2$ is the kinetic energy of a molecule of mass $m$, and $k_B T = \la m  v^2 \ra/2$ is the average kinetic energy of all the molecules.   Note that this system has a single energy scale $ k_B T$.  Also, the system is in equilibrium, and it obeys principle of detailed energy balance.  As a result, there is no energy  transfer from one region to another, both in real and Fourier space.  

Now we are ready to describe the equilibrium model of wealth distribution~\cite{Saha:book:Heat,Chakrabarti:book:Econophysics}.  In the past, several researchers have shown connections between economic systems and equilibrium thermodynamics (e.g., a gas reservoir described above)~\cite{Saha:book:Heat}. The individuals or economic entities are analogous to the gas molecules, and wealth to the kinetic energy of the molecules.  Refer to Table~\ref{tab:mkv:1} for a detailed comparison.
\begin{table}
\caption{Analogies between an equilibrium economic model and a thermodynamic system}
\label{tab:mkv:1}       
\begin{tabular}{p{5cm}p{4cm}}
\hline \noalign {\smallskip}
Thermodynamics & Ecomomics  \\
Thermodynamic system & Economy \\
Gas molecules 		& Economic entities (individuals) \\
Individual kinetic energy 		& Individual wealth \\
Collisions 			& Economic Interactions \\
Average kinetic energy		& Average wealth \\
\end{tabular}
\end{table}
Using this analogy, researchers deduced that the wealth distribution $P(W)$ in an economy  follows Maxwell or Gibbs distribution:
\be
P(W) = \exp(-W/ \la W \ra),
\ee
 where $\la W \ra$ is the average wealth of an individual in the economic system.  More refined models yield log-normal distribution~\cite{Chakrabarti:book:Econophysics}.

\section{Multiscale model of Turbulence}
\label{sec:3}

\begin{figure}[b]
\includegraphics[scale=.50]{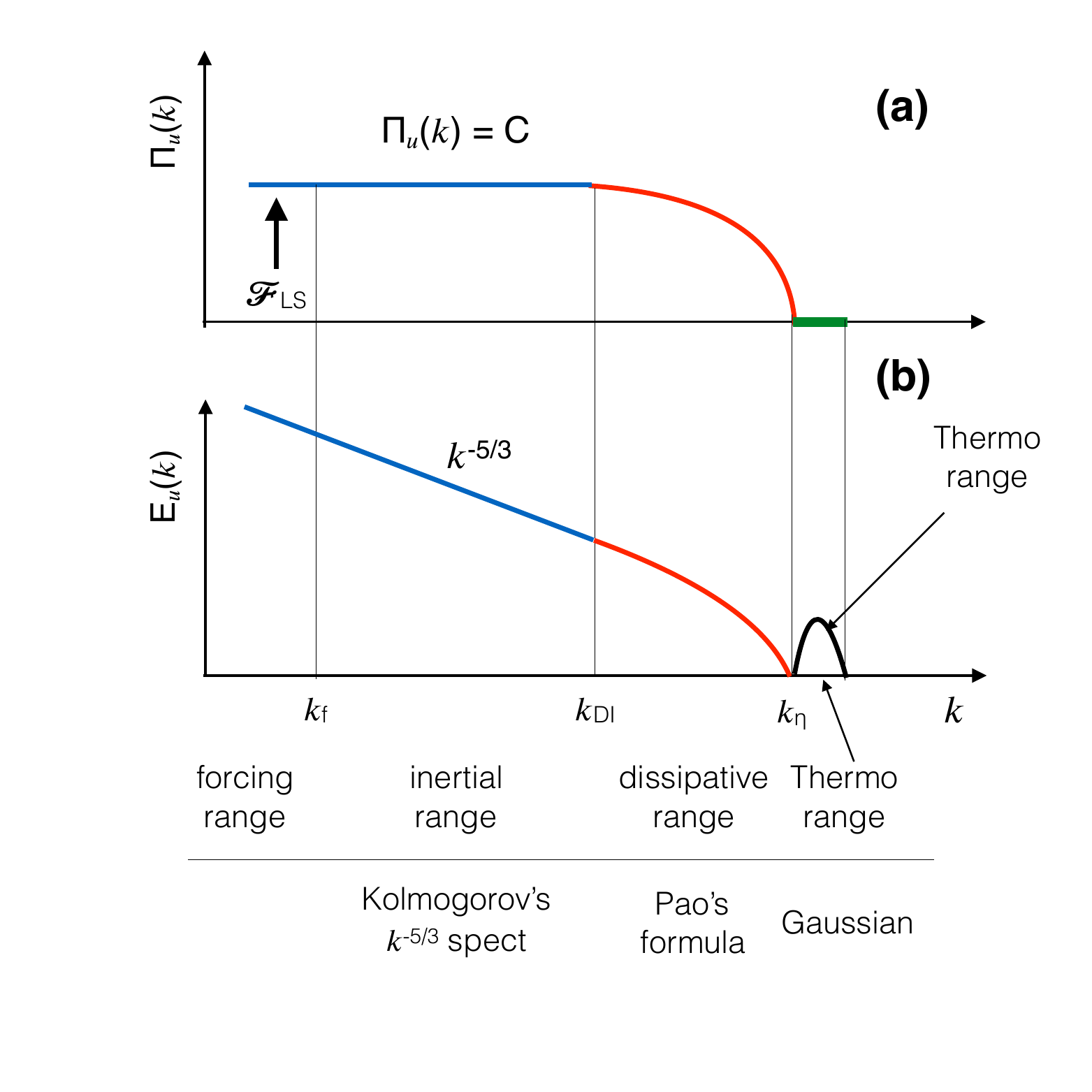}
\caption{Kolmogorov's picture of hydrodynamic turbulence.  The flow is forced at large scale with an energy injection rate of $\mathcal{F}_\mathrm{LS}$.  (a) The energy flux is constant in the inertial range, and it decays in the dissipative range.  Flux is zero in the thermodynamic range. (b) The energy spectrum exhibits $k^{-5/3}$ spectrum in the inertial range.  In the thermodynamic range, the molecules of the fluid exhibit Maxwellian distribution (black curve).}
\label{fig:mkv:hydro}       
\end{figure}
Many nonequilibrium systems have properties very different from that of the gas reservoir described above~\cite{Ma:book:StatMech}.  For example, consider a turbulent fluid stirred at large length scaless.  The kinetic energy at large scales cascades to intermediate scale, and then to small scales.  The kinetic energy flux $\Pi_u$ is constant in the inertial range, and then it decreases in the dissipation range.  Due to the constant energy cascade, principle detailed balance is broken in such a system. The energy flux is zero at microscopic scale where we expect thermodynamic principles to hold.  This is Kolmogorov's picture of hydrodynamic turbulence~\cite{Kolmogorov:DANS1941Dissipation,Kolmogorov:DANS1941Structure,Lesieur:book:Turbulence,Verma:book:BDF}. See Fig.~\ref{fig:mkv:hydro} for an illustration of the energy flux and energy spectrum.

The energy spectrum $E_u(k)$ of a turbulent flow has been derived using dimension analysis.  Using 
\be
[E_u(k)] = [E_u/k] = [L^3/T^2];~~[\Pi_u] = [E_u/T] = [L^2/T^3];~~[k]=[L]^{-1},
\ee
 we derive  the following formula for the kinetic energy spectrum:
\be
E_u(k) =  K_\mathrm{Ko} \epsilon_u^{2/3} k^{-5/3},
\label{eq:Kolm_Ek}
\ee
where $K_\mathrm{Ko}$ is  Kolmogorov's constant, and $\epsilon_u$ is the kinetic energy dissipation rate.  Pao~\cite{Pao:PF1968} extended the above formula to the dissipation range, and obtained
\bea
\Pi_u(k)  & =  & \epsilon_u \exp{\left(- \frac{3}{2} K_\mathrm{Ko}  (k/k_d)^{4/3}\right)}, \label{eq:hydro_3d:Pao_Pik}\\
E_u(k)  & = & K_\mathrm{Ko}  \epsilon_u^{2/3} k^{-5/3}  \exp{\left(- \frac{3}{2} K_\mathrm{Ko}  (k/k_d)^{4/3}\right)}, \label{eq:hydro_3d:Pao_Ek}
\eea 
where $k_d$ is  Kolmogorov's wavenumber~\cite{Leslie:book}.  The fluid kinetic energy vanishes beyond the dissipation range, i.e., for $k >k_d$.  The above function describes the inertial and dissipative ranges (blue and red curves of Fig.~\ref{fig:mkv:hydro}) quite well.   We expect thermodynamic ideas to work beyond this scale.  The energy of the molecules would follow Maxwell's or Gibbs' distribution, as shown by the black curve of Fig.~\ref{fig:mkv:hydro}(b).

Shell model is a popular model of hydrodynamic turbulence. In one version of the shell model, called GOY shell model of turbulence, 
\be
\frac{d}{dt} u_n+\nu k_{n}^{2}u_n=-i(a_{1}k_{n}u_{n+1}^{*}u_{n+2}^{*}+a_{2}k_{n-1}u_{n+1}^{*}u_{n-1}^{*}+a_{3}k_{n-2}u_{n-1}^{*}u_{n-2}^{*}),\label{eq:fluid}
\ee
where $u_n$ is a complex number representing the velocity field at length scale $k_n = 2^n$, $a_1, a_2, a_3$ are constants, and $\nu$ is the kinematic viscosity~\cite{Ditlevsen:book}.  In the shell model, the kinetic energy follows 
\be
E(k_n) = \frac{|u_n|^2}{2 k_n} \sim k_n^{-5/3},
\ee
in accordance with Kolmogorov's theory of turbulence.   Our finance model has a similar form as the above shell model, as we will describe in the next section. 

\section{A model of hierarchical financial entities}
\label{sec:mkv:finance_model}

\begin{figure}[b]
\begin{center}
\includegraphics[scale=.50]{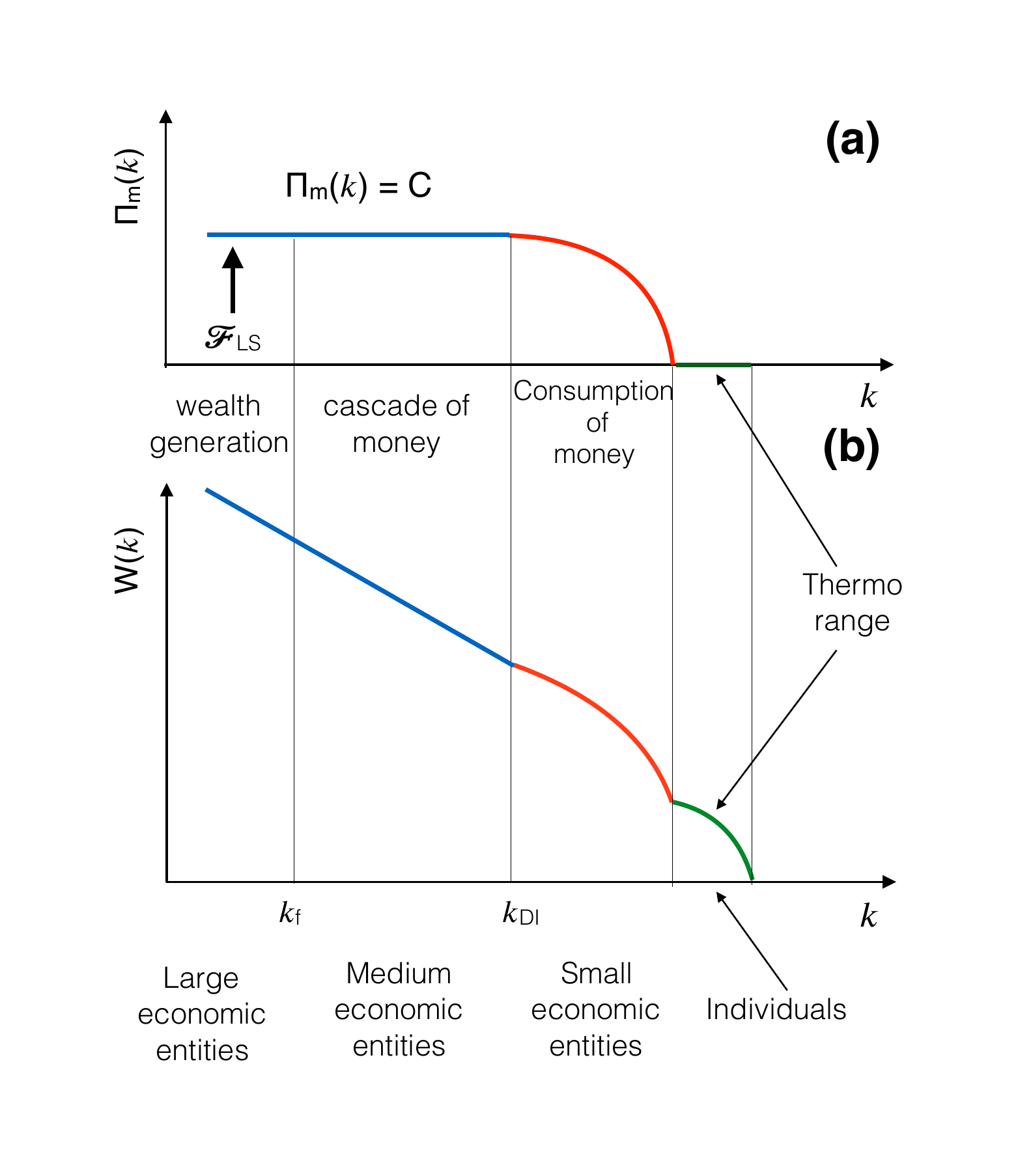}
\end{center}
\caption{In a hierarchical finance model, (a) the flux of money $\Pi_m(k)$, and (b) the wealth distribution of financial entities.  The large and medium economic entities have constant money flux and power law distribution of wealth.  Small economic entities exhibit exponential distribution, while the thermodynamic range exhibits Maxwellian wealth distribution and zero money flux. }
\label{fig:mkv:finance}       
\end{figure}
 We construct a model for a hierarchical finance system in a similar lines as Kolmogorov's picture for hydrodynamic turbulence.   In this model, we assume that the wealth is generated at the largest scale, and then it  flows from larger financial structures to smaller structures in a steady manner.   We also assume that the financial  entities of similar sizes interact with each other. This is similar to the local interactions in turbulence.  In addition, in the absence of financial pilferage, we expect that the cascade of money from large structures to smaller structures to be a constant.  This is same as the assumption of constant energy cascade in hydrodynamic turbulence. 
 See Fig.~\ref{fig:mkv:finance} for an illustration, and  Table~\ref{tab:mkv:finance_model} for listing of similarities between a turbulent system and a hierarchical financial system.
\begin{table}
\caption{Analogies between turbulence and hierarchical financial system}
\label{tab:mkv:finance_model}       
\begin{tabular}{p{5cm}p{5cm}}
\hline \noalign {\smallskip}
Turbulence  & financial system  \\
Fluid structures 		& Financial entities \\
Multiscale 		&	Multiscale \\
kinetic energy of a structure 	& Wealth of a financial entity \\
Constant energy flux	 	&  Constant money supply \\
Power law $E_u(k)$ at intermediate scales & Power law for large income entities \\
Exponential  $E_u(k)$ at small scales & Expect similar scaling \\
Random motion beyond $k_d$ & Gibbs distribution at individual scale \\
\end{tabular}
\end{table}

We place these financial entities in a two-dimensional wavenumber grid~\footnote{Dimensionality of a hierarchical financial system is an undetermined parameter.  Here we choose $d=2$ using an observation that these structures reside on the surface of the Earth.}.   Let us denote the financial asset of a financial entity at the wavenumber ${\bf k}$  as $W({\bf k})$.     The number of mesh points on a 2D disc of radius $k$ is 
\be
n(k) = 2 \pi k.
\label{eq:mkv:n_k}
\ee
We solve for the wealth distribution as a function of $n$.  To illustrate, there are fewer financial entities at small $k$,  corresponding to financial giants (e.g. Google and Apple of today).  Large number of modes at large $k$ correspond to small units like small  companies or individuals.

Motivated by the shell model of turbulence, we construct the following model for the hierarchical financial entities:
\be
\frac{d W_k}{dt} = a k^\alpha W_{k-1} W_{k+1} - b k^\beta W_k + Q_{k,1},
\label{eq:mkv:W_k_dot}
\ee
where $a,b, \alpha$ and $\beta$ are constants, and $W_k$ is analogous to the shell spectrum in turbulence. Hence, 
\be
W_k = 2\pi k W({\bf k}).
\label{eq:mkv:W_k_W_bfk}
\ee In Eq.~(\ref{eq:mkv:W_k_dot}), the first term in the RHS represents  the interactions among financial entities at scales $k, k-1$ and $k+1$, while the second term represents financial losses at scale $k$ (e.g., recurring expenses, e.g., electricity bills).  The third term $Q_{k,1}$ represents the wealth generation at the largest scale, $k=1$.  

This is a very simple model because it ignores nonlocal interactions, as well as other complex things like loans, savings, banks, generation of wealth at the intermediate and small scales, etc.  Further, we assume a steady state in which money flows from larger structures to smaller structures.  The wealth is finally consumed at the smallest structures of the system.  

\begin{figure}[b]
 \begin{center}
\includegraphics[scale=.50]{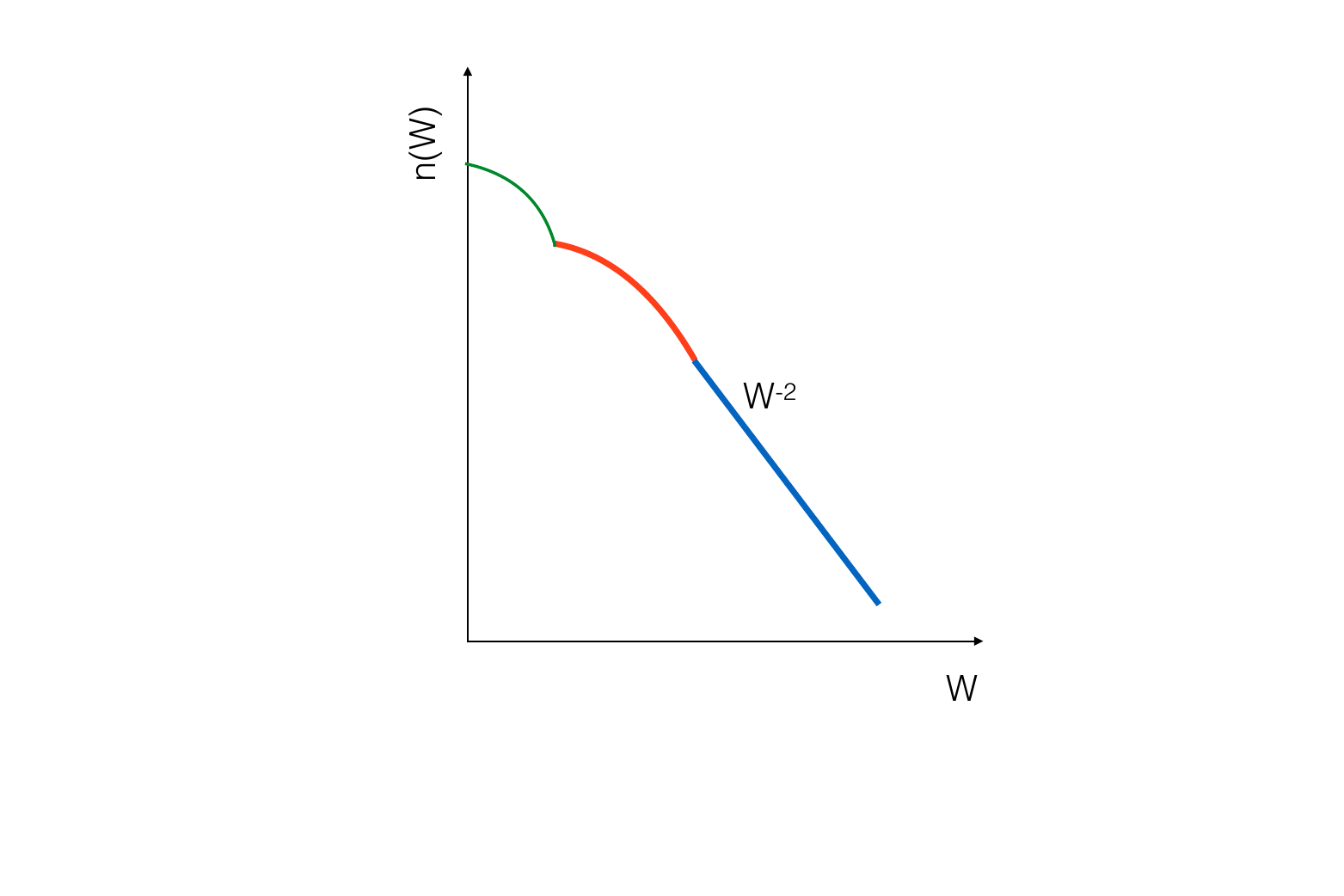}
 \end{center}
\caption{Plot of wealth distribution $n(W)$ vs. $W$ that indicates number of financial entities with wealth $W$.}
\label{fig:mkv:nW_vs_W}       
\end{figure}

First, we focus on the large and intermediate scale where we expect  a power law scaling.  We also assume that the financial losses at these scales are negligible. Under a steady state,
\be
\Pi = \frac{d W_k}{dt} \sim  k^\alpha W_{k}^2,
\label{eq:mkv:Pi}
\ee
where $\Pi$ is the cascade of money.  We invert Eq.~(\ref{eq:mkv:Pi}) that yields
\be
W_k \sim \Pi^{1/2}  k^{-\alpha/2}.
\ee
Now using Eqs.~(\ref{eq:mkv:n_k}, \ref{eq:mkv:W_k_W_bfk}), we obtain
\be
n(W) \sim \Pi^{\frac{1}{\alpha+2}} W^{-\frac{2}{\alpha+2}},
\label{eq:mkv:n(W)}
\ee
where we write $W({\bf k})$ as $W$.  The above formula yields the number of financial entities $n(W)$ with wealth $W$. Clearly, $\alpha = -1$ gives
\be
n(W) \sim \Pi W^{-2},
\ee
which is similar to the Pareto's law for the wealth distribution~\cite{Pareto:book, Chakrabarti:book:Econophysics} of the large financial entities.  Note however that the exponent depends quite crucially on the choice of $\alpha$.  In Fig.~\ref{fig:mkv:nW_vs_W}, we exhibit the inverted form of Fig.~\ref{eq:mkv:n_k}(b), or the plot of wealth distribution $n(W)$ vs. $W$.

The wealth cascades down to smaller scales, and it finally gets consumed at the dissipation scales (individual level).  It could be in the form of consumption of food and basic needs.  Following the popular equilibrium model~\cite{Saha:book:Heat}, the wealth distribution at this scale follows Maxwellian or Gibbs distribution.  We also expect an income group between the power law regime and the Gibbs distribution.  This regime may follow a law similar to that Pao's model for turbulence, which was discussed in Section 3.

Several cautionary remarks are in order.  Our model describes the wealth of financial entities.  Pareto's law however is stated for individual incomes.  In free market, a financial entity is essentially owned or controlled by several individuals or a group of individuals.  Therefore, it is reasonable to assume that the wealth distribution of financial entities also reflects the wealth or income distribution of individuals.  Also, a large financial entity contains smaller entities, thus  forming hierarchical structures. 

A corollary to the above model is as follows.   Let us consider finance distribution in a country.  The central government transfers resources to various states who distributes it to lower levels in a hierarchical manner, e.g., 
\be
\mathrm{states} \rightarrow \mathrm{district} \rightarrow \mathrm{village}.
\ee 
Following the same line of arguments as before, we deduce that the financial resources at hierarchical level must be a power law.  If there is no corruption, then the money supply at different levels is constant.

The above model is very simple.  It ignores many important ingredients such as savings, stocks, banking, pilferage of wealth, nonlocal interactions, etc.  This model however has certain novelty.  It emphasises on multiscale nature of financial systems, cascade of money at different scales, and nonequilibrium nature of the financial system.

\section{Discussions and Conclusion}

Our finance model, though simple, captures multiscale economic transaction among financial entities and explains coexistence of power law and Maxwellian distribution for the wealth of these entities~\cite{Chakrabarti:book:Econophysics}.    The model has other predictions as well. Note that the model has a free parameter $\alpha$.   The present multiscale model has many assumptions that may not hold for real financial system.  For example, we need to include savings, banking, variable energy flux, etc. in a more refined model.  In addition, we need to contrast the present model with the existing models, some of which are described in \cite{Chakrabarti:book:Econophysics,Chakraborti:EPJB2000,Patriarca:PRE2004,Chakraborti:PRL2009}

A small financial system without hierarchy may exhibit detailed balance and Maxwellian distribution for the wealth distribution.  As soon as a financial system becomes sufficiently large and it follows a free-market economy, we expect wealth inequalities to develop based on individual abilities and ambitions.  Such a system will exhibit power law distribution.   Strong economic regulations  may suppress the inequality and make the power law shallower.  

We believe that the finance model presented here shed important insights into financial systems. Yet, it is a preliminary model and it needs more work and refinements.

\acknowledgements
I am thankful Anirban Chakraborty, Supratik Banerjee, Andre Sukhanovskii, Rodion Stepanov, Franck Plunian, Abhishek Kumar, and Kiran Sharma for very useful discussions and ideas.  


\end{document}